%% file: template.tex
\documentclass[preprint]{vgtc}               

\graphicspath{{content/papers/sphericolor/figures/}{pictures/}{images/}{./}} 

\usepackage{times}                     

\usepackage{tabu}                      
\usepackage{booktabs}                  
\usepackage{lipsum}                    
\usepackage{mwe}                       

\usepackage{mathptmx}                  

\onlineid{1182}

\vgtccategory{Research}

\vgtcinsertpkg

\input{sections/_preamble}

\title{\papertitle}

\author{Julius Rauscher\thanks{e-mail: {firstname.lastname}@uni-konstanz.de}\\ %
        \scriptsize University of Konstanz %
\and Johannes Fuchs\\ %
     \scriptsize University of Konstanz %
 \and Daniel A. Keim\\ %
     \scriptsize University of Konstanz %
      \and Frederik L. Dennig\\ %
     \scriptsize University of Konstanz }

\teaser{

  \includegraphics[width=\linewidth]{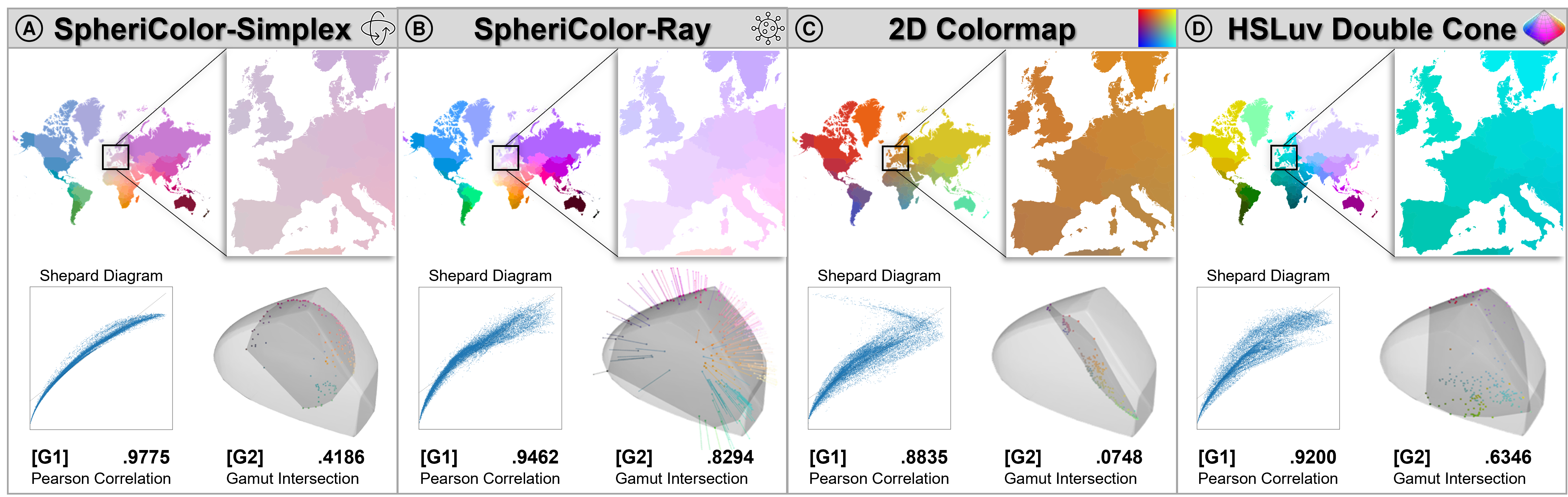}
 \centering
  \caption{
   Comparison of approaches to generate colors from spherical geographic inputs (world countries). a) \mapSimplex scales and rotates a 3D MDS embedding to fit inside the sRGB gamut. This preserves the input distances (highest Pearson correlation), but does not fully exploit the available colorspace.  b) \mapRay projects the embeddings on the gamut hull along rays directed at the gamut centroid, yielding a lower distance preservation but a larger colorspace exploitation (Gamut intersection). 
   c) 2D colormaps suffer from large color distances for small geodesic distances when crossing the antimeridian. d) The HSLuv double cone provides adequate, but inferior results when compared to \mapname approaches. 
   }
\label{sphericolor:fig:teaser}
}

\abstract{
    \input{sections/00_abstract}
} 


\begin{document}



\maketitle
 
\input{sections/01_introduction}

\input{sections/02_related_work}

\input{sections/03_approach}

\input{sections/04-quantiative-eval}

\input{sections/06_discussion}

\input{sections/07_conclusion}

\section*{Acknowledgement}
This work was funded by the Deutsche Forschungsgemeinschaft (DFG) -- TRR 161 (Project A03) -- 251654672.

\bibliographystyle{abbrv-doi}

\bibliography{references/refs}       

\end{document}


\maketitle

\renewcommand{\thefigure}{A.\arabic{figure}}
\setcounter{figure}{0}

\newcommand{\ColormapSummary}[2]{%
\begin{figure*}[t]
    \centering
    {\Huge\textbf{#1}}\par\vspace{1em}

    \begin{tabularx}{\linewidth}{X !{\vrule width 0.1pt} X}

    \begin{minipage}[t]{\linewidth}
        \centering
        {\Large\textbf{Countries}}\par\vspace{0.5em}
        \vspace{2em}
        
        \begin{subfigure}[t]{.7\linewidth}
            \centering
            \includegraphics[width=\linewidth]{content/papers/sphericolor/figures/approach/eval/countries_#2_shepard.png}
            \caption{Shepard Diagram}
        \end{subfigure}

        \begin{subfigure}[t]{\linewidth}
            \centering
            \includegraphics[width=\linewidth]{content/papers/sphericolor/figures/approach/eval/countries_#2_3d.png}
            \caption{CAM16-UCS Embedding}
        \end{subfigure}

        \begin{subfigure}[t]{\linewidth}
            \centering
            \includegraphics[height=15em]{content/papers/sphericolor/figures/approach/eval/countries_#2_map.png}
            \caption{Geospatial View}
        \end{subfigure}

    \end{minipage}
    &
    \begin{minipage}[t]{\linewidth}
        \centering
        {\Large\textbf{Samples}}\par\vspace{0.5em}
        \vspace{2em}

        \begin{subfigure}[t]{.7\linewidth}
            \centering
            \includegraphics[width=\linewidth]{content/papers/sphericolor/figures/approach/eval/samples_#2_shepard.png}
            \caption{Shepard Diagram}
        \end{subfigure}

        \begin{subfigure}[t]{\linewidth}
            \centering
            \includegraphics[width=\linewidth]{content/papers/sphericolor/figures/approach/eval/samples_#2_3d.png}
            \caption{CAM16-UCS Embedding}
        \end{subfigure}

        \begin{subfigure}[t]{\linewidth}
            \centering
            \includegraphics[height=15em]{content/papers/sphericolor/figures/approach/eval/samples_#2_map.png}
            \caption{Geospatial View (using Voronoi tessellation)}
        \end{subfigure}

    \end{minipage}

    \end{tabularx}

    \caption{Overview for colormap configuration #1.}
\end{figure*}
}


\section{Mathematical Formulas}

The following section provides additional information about the proposed optimization algorithms. The full code for replication of the method, as well as generating the data and figures for the quantitative evaluation, is available on \href{https://osf.io/h84kn/overview?view\_only=53355eee9b6741d3bcc0fc2285bbd112}{OSF}. 

\subsection{\mapSimplex}
The following objective function is minimized using Nelder-Mead simplex optimization to determine the transformation of the MDS embedding:
\begin{equation}
f(\mathbf{T}) = w_{\text{stress}} \cdot \mathcal{L}_{\text{dist}}(\mathbf{T}) + w_{\text{gamut}} \cdot \mathcal{L}_{\text{hull}}(\mathbf{T}) - w_{\text{expand}} \cdot s(\mathbf{T})
\end{equation}
where $\mathbf{T} = (t_{J^*},t_{a^*},t_{b^*},\theta, s)$ represents a rigid transformation with three translation components (corresponding to the $J^*$,$a^*$,$b^*$ components of the CAM16-UCS), one rotation angle, and a uniform scale factor.
$\mathcal{L}_{\text{dist}}$  measures distance preservation as the mean squared stress between pairwise distances in the transformed embedding and the input geodesic distance matrix.
$\mathcal{L}_{\text{hull}}$ penalizes out-of-gamut points as the sum of signed distances to the nearest convex hull plane of the sRGB gamut, and $s$ rewards expansion.

As weights, we determined:

\noindent $w_{\text{stress}} = 0.1 $, 
$w_{\text{gamut}} = 1.0 $, and $w_{\text{expand}} = 0.1 $

A higher weighting on the penalization of out-of-gamut points $\mathcal{L}_{\text{hull}}$ ensures that the embedding is constrained within the sRGB gamut.

The optimization is initialized with the translation set to the gamut centroid, rotation and scale set to $0$, and initial simplex step sizes of $[150,150,150]$ for each translation component and $1$ for rotation and scale.

\subsection{\mapRay}
The ray-based optimization expects a 3D embedding that is centered around the centroid of the sRGB gamut. 
Each point $\mathbf{p}_i$ onto the gamut surface along rays from the centroid $\mathbf{c}$ using the following equation:
\begin{equation}
\mathbf{p}_i' = \mathbf{c} + t_i \cdot \frac{\mathbf{p}_i - \mathbf{c}}{\|\mathbf{p}_i - \mathbf{c}\|}
\end{equation}
where $t_i$ is the ray-triangle intersection distance to the gamut hull.

\subsection{Colorspace Exploitation -- Gamut Intersect}

To approximate the colorspace exploitation, we determine the volume of the colormap in CAM16-UCS and divide this by the volume of the sRGB gamut to normalize it to a score between $[0,1]$, where $1$ represents maximum exploitation (the colormap completely fills the sRGB gamut).

To ensure we measure only displayable colors, we first intersect the colormap's convex hull with the sRGB gamut boundary before computing the volume.

\begin{equation}
\text{Gamut Intersect} = \frac{V(\text{hull}(\mathcal{C}) \cap \Gamma_{\text{sRGB}})}{V(\Gamma_{\text{sRGB}})}
\end{equation}
where $\mathcal{C} = \{\mathbf{c}_i\}$ is the set of colormap colors in CAM16-UCS, $\text{hull}(\mathcal{C})$ is their convex hull, $\Gamma_{\text{sRGB}}$ is the sRGB gamut polyhedron, and $V(\cdot)$ denotes volume.

\onecolumn

\ColormapSummary{\mapSimplex}{sphericolor-simplex}
\ColormapSummary{\mapRay}{sphericolor-ray}
\ColormapSummary{2D Colormap}{2dcolormap}
\ColormapSummary{HSLuv Double Cone}{hsl}
\ColormapSummary{HSL Double Cone}{hsll}


%% file: sections/_preamble.tex
\usepackage[english]{babel}
\usepackage[svgnames]{xcolor}
\usepackage{wrapfig}
\usepackage{graphicx}
\usepackage{stfloats}
\usepackage{caption}
\usepackage{subcaption}
\usepackage{dsfont}
\usepackage{enumitem}
\usepackage{siunitx}
\usepackage[T1]{fontenc}
\usepackage{booktabs}
\usepackage{multicol}
\usepackage{multirow}
\usepackage{xspace}
\addto\extrasenglish{

}

\usepackage{tabularx}

\usepackage{microtype}

\newcommand\figVSpace{-1.5em}

\newcommand{\subhead}[1]{\vspace{4pt} \noindent \textbf{#1 --}}

\newlength{\blen}
\definecolor{color_distance}{HTML}{E0AD41}
\definecolor{color_value}{HTML}{6278A1}

\newcommand{\compresslist}{
	\setlength{\itemsep}{1pt}
	\setlength{\parskip}{0pt}
	\setlength{\parsep}{0pt}	
}

\newcommand{\revise}[1]{\textcolor{black}{#1}}
\newcommand{\revision}[1]{\textcolor{black}{#1}}

\newcommand{\mapname}[0]{SpheriColor\xspace}
\newcommand{\countryds}[0]{\emph{Countries}\xspace}
\newcommand{\sampleds}[0]{\emph{Samples}\xspace}

\newcommand{\mapSimplex}[0]{\mapname-Simplex\xspace}
\newcommand{\mapRay}[0]{\mapname-Ray\xspace}

\newcommand{\papertitle}{\mapname: Colormaps for Spherical Geospatial Input Topographies}

\newcommand{\GoalA}{\textbf{[G1]}\xspace}
\newcommand{\GoalB}{\textbf{[G2]}\xspace}

%% file: sections/00_abstract.tex
Multivariate geospatial data visualization often relies on multiple coordinated views, where color can be used to either link views or encode data attributes.
Encoding spatial locations through color can reveal patterns in non-spatial visualizations, yet most applications of colormaps focus on high-dimensional attribute encodings instead.
While 2D colormaps have been studied extensively, color encodings designed for spherical geospatial data have received much less attention, even though all locations lie on a sphere.
To address this, we propose \mapname, a colormap generation approach for mapping geospatial references guided by the design goals of distance preservation and colorspace exploitation.
We project a geospatial distance function into a perceptually linear colorspace, followed by two different gamut-constrained optimization strategies.
We perform a quantitative evaluation using both real-world and synthetic datasets, demonstrating superior performance over 2D colormaps and HSLuv double cone encodings. The simplex-based optimization excels at distance preservation, whereas the ray-based approach provides better colorspace exploitation.

%% file: sections/01_introduction.tex
\section{Introduction}
\label{sphericolor:sec:introduction}
The availability of large multivariate datasets with spatial references has grown significantly across various fields, such as  economics~\cite{WorldDevelopmentIndicators:worldbank:2014} or demographics~\cite{Moosavi:hdspatialcolor:2017}. 
Visual analysis of such data commonly combines geographic map visualizations with attribute views such as Parallel Coordinate Plots (PCP)~\cite{opach:ChoroplethMapsLinked:2014,liu:MapinParallelCoordinatesPlot:2023} or scatterplots derived from dimensionality reduction techniques~\cite{visstamp:guo:06, mazher:hdspatialcolor:2020, mvst:lu:2023}.
Color can be employed to assess similarities on either the \emph{spatial} or \emph{non-spatial} component, where existing approaches tend to focus on encoding the high-dimensional attributes into colorspaces through Self-organizing Maps (SOM)~\cite{kaski:ColoringThatReveals:2001, Moosavi:hdspatialcolor:2017}, Multidimensional Scaling (MDS)~\cite{mittelstadt:RevisitingPerceptuallyOptimized:2014}, or Uniform Manifold Approximation and Projection (UMAP)~\cite{koutrouli:UCIE:2022}.
Encoding geospatial data with color has received less attention, despite its relevance for tasks in which spatial relationships are the analysis target, although \revision{such encodings can visually reveal spatial patterns in non-spatial visualizations such as PCPs or scatterplots.}
\revise{
As geography is conventionally mapped to position, encoding it through color instead frees position as a visual variable for additional attributes. 
}

Guidelines for designing color mappings exist for continuous, 1D scalar ranges~\cite{colormapTaxonomy:bujack:2018} or 2D colormaps~\cite{survey2dcolormap:bernard:15}, highlighting distance preservation and colorspace exploitation for quality assessment, where a trade-off is unavoidable. 
\revision{Preserving distances ensures that perceived color distances reflect data distances, while exploiting the colorspace maximizes color discriminability.
Such guidelines, however, have not been extended to spherical geospatial inputs.}
While map projections can be leveraged to rely on the existing literature on 2D colormaps~\cite{assistedDescriptorSelection:bremm:14, mittelstadt:RevisitingPerceptuallyOptimized:2014, survey2dcolormap:bernard:15}, these methods have shortcomings when applied to the entire spherical topography of the globe, such as extreme color discontinuities along the antimeridian where the globe is typically cut to unfold the sphere to a plane (see~\autoref{sphericolor:fig:teaser}c).


To preserve geodesic distances in color space, perceptually uniform color spaces (UCS) can be used, but standard displays can only reproduce a limited subset of these color spaces. 
Out-of-gamut colors are clipped or remapped, introducing distortions that impair distance preservation~\cite{colortasks:tominski:08, colorpalettes:wijffelaars:2008, diverging:moreland:2009}. Optimization methods that constrain colormap embeddings to device gamuts have been proposed~\cite{koutrouli:UCIE:2022}, but have not been quantitatively evaluated with respect to the trade-off between distance preservation and colorspace exploitation.
In this paper, we contribute:

\begin{itemize}[left=0pt]
\compresslist
\item \mapname, a method for creating perceptually linear mappings from spherically bound geographic references to colors \href{https://osf.io/h84kn/overview}{(OSF)}.
\item A quantitative comparison to other methods in regard to the imposed design goals derived from existing literature. 
\end{itemize}

%% file: sections/02_related_work.tex
\section{Related Work}
\label{sphericolor:sec:relatedwork}


\subhead{Multivariate Spatial Visualization}
Glyph maps, either gridded~\cite{slingsby:gridmapcovid:2023} or entity-centered~\cite{mcnabb:MultivariateMapsGlyphPlacement:2019}, are a commonly utilized superimposing visualization technique for multivariate data with spatial context, but suffer from visual overload as dimensionality increases~\cite{laksono:gridglyphmap:2024}.
Early work by Andrienko and Andrienko~\cite{andrienkoExploringSpatialData2001} combined a PCP with choropleth maps for multivariate exploration. Subsequent user studies~\cite{opach:ChoroplethMapsLinked:2014, elmeseery:GeoCoordinatedParallelCoordinates:2018} show that such juxtaposed layouts coupled with an attribute-based colorscheme support identifying spatial patterns in the map. Liu et al.~\cite{liu:MapinParallelCoordinatesPlot:2023} further integrate both views by embedding the map within the PCP using connecting polylines.

Concerning the usage of color, Kaski and Venna~\cite{kaski:ColoringThatReveals:2001} project high-dimensional data with a SOM to 2D and derive a colormap from a CIELab cross section for choropleth mapping. Moosavi~\cite{Moosavi:hdspatialcolor:2017} applies a 1D SOM-based encoding mapped to a colorscale with an accompanying radial diagram.
Mazher~\cite{mazher:hdspatialcolor:2020} juxtaposes a scatterplot and a rasterized map, using a coloring obtained from different dimensionality reduction methods and a 2D colormap. 
More comprehensive visual analytics systems~\cite{visstamp:guo:06, mvst:lu:2023} integrate multiple coordinated views (scatterplots, matrix views, or storyline views) for multivariate spatiotemporal analysis.
Existing approaches for spatially-referenced multivariate data use color only for non-spatial attributes. To the best of our knowledge, no sophisticated method maps geographic data to colormaps.

\subhead{Color Encodings for Multivariate Data}
Zhou and Hansen~\cite{zhouHansen:colormapsurvey:2016} survey colormap design strategies, classifying them as procedural, user-study based, rule-based, or data-driven. For 1D continuous colormaps, Bujack et al.~\cite{colormapTaxonomy:bujack:2018} review design rules, extended by automated improvements from Nardini et al.~\cite{colormapImprovements:nardini:2021}. Bernard et al.~\cite{survey2dcolormap:bernard:15} survey 2D colormaps, while Steiger et al.~\cite{explorative2dcolormaps:steiger:15} emphasize task-dependent selection and provide interactive guidance. 
Waldin et al.~\cite{personalized2dcolor:waldin:16} introduce personalized 2D colormaps for users with color deficiencies. Mittelstädt et al.~\cite{mittelstadt:RevisitingPerceptuallyOptimized:2014} stress perceptual linearity in 2D colormaps and employ dimensionality reduction to encode high-dimensional data. Cheng et al.~\cite{colormapnd:cheng:19} interactively map multivariate data onto a circular HCL representation, and Koutrouli et al.~\cite{koutrouli:UCIE:2022} generate perceptually linear colormaps by embedding UMAP-based 3D projections into CIELAB.
In Diffusion Tensor Imaging, spherical mappings to color have been explored for encoding fiber orientations using absolute value RGB~\cite{absoluteDTIColormap:pajevic:1999}, or Boy's surface immersion~\cite{boysDTIcolormap:demiralp:2009}, though these address line orientations rather than geographic positions.

While several approaches for colormap generation have been proposed, none have considered geographic positional inputs from spherical topographies to the best of our knowledge, and no empirical quantification has been carried out to assess the effectiveness of such approaches.





%% file: sections/03_approach.tex
\section{Spherical Spatial Colormaps}
\label{sphericolor:sec:approach}
We provide design goals for the \revision{data-driven} generation of a colormap encoding \revision{a fixed set of} spherical geospatial references, outline naïve approaches and their shortcomings, and show how our novel approach fulfills the imposed criteria.

\subsection{Design Goals}

Previous work on colormap generation has often emphasized the tradeoff between perceptual linearity and color space exploitation~\cite{assistedDescriptorSelection:bremm:14,mittelstadt:RevisitingPerceptuallyOptimized:2014, survey2dcolormap:bernard:15}.
Bujack et al. \cite{colormapTaxonomy:bujack:2018} present a theoretical framework for assessing continuous colormaps, providing a summary of design rules distilled from the existing literature, providing mathematical formulas for the concepts of \emph{Uniformity}, \emph{Discriminative power}, and \emph{Order}.
For spherical inputs, a natural total ordering does not exist, which is why we do not consider \emph{Order preservation} a design goal.
%
Hence, we derive our design goals and describe them as follows:

\noindent\textbf{\GoalA Distance Preservation --}
Corresponding to \emph{Uniformity}, objects close in space shall correspond to similar colors, while objects far apart shall be proportionally different in their perceived color.\\
\textbf{\GoalB Colorspace Exploitation --}
The colormap shall optimize color usage to maximize \emph{discriminative power}. The output device imposes the final constraint, which we define as the sRGB gamut to ensure standard devices can display all colors.



\subsection{Naïve Approaches}

\subhead{2D Colormap}
The widespread use of 2D colormaps~\cite{survey2dcolormap:bernard:15, explorative2dcolormaps:steiger:15} and 2D geographical maps suggests a simple superimposition. \autoref{sphericolor:fig:teaser}c illustrates this using the \emph{Cube Diagonal Cut B-C-Y-R} colormap~\cite{survey2dcolormap:bernard:15} with an overlaid Mercator projection. While this leverages well-studied 2D colormaps, it inherits projection-induced distortions when mapping a spherical surface to a plane. In particular, cylindrical projections cut the sphere at a given longitude, commonly the antimeridian, violating \GoalA\ by placing nearby regions on opposite sides of the colormap.
According to Koutrouli et al.~\cite{koutrouli:UCIE:2022} human color perception is inherently three-dimensional, and Mittelstädt et al.~\cite{mittelstadt:RevisitingPerceptuallyOptimized:2014} further point out that perceptual uniformity for 2D colormaps can only be achieved at the cost of unsaturated and less discriminative colors, limiting the colorspace exploitation \GoalB.

\subhead{HSLuv Double Cone}
The HSL double cone represents color with hue as a circular dimension, saturation increasing radially, and lightness along the vertical axis, forming two cones joined at their tips. 
HSLuv~\cite{boronine:hsluv:2025}, an extension of CIELUV, can be used to reduce perceptual nonlinearities of HSL~\cite{colormapnd:cheng:19}. Mapping longitude to hue and latitude to lightness yields a continuous color cycle over the sphere, transitioning from dark colors at the south pole to bright ones at the north pole (see~\autoref{sphericolor:fig:teaser}d). 
Unlike planar 2D colormaps, this representation avoids projection-induced discontinuities as the cyclical hue wraps smoothly around the globe, and mapping to the fully saturated outer hull fully exploits the available colorspace~\GoalB. However, the double cone only approximates a sphere, and the deformation by HSLuv further distorts the original geometry, leading to deviations between input distances and output color differences, violating \GoalA.



\subsection{\mapname}

Following prior work on mapping high-dimensional data to colorspaces~\cite{koutrouli:UCIE:2022}, we compute a distance matrix, project it into three dimensions, followed by an optimization step.
While spherical coordinates can be directly converted to 3D Euclidean space, we employ a distance-based approach as it offers greater flexibility.
This allows the integration of more complex geographical distances such as travel times, refined polygon similarity measures~\cite{1dorder:rauscher:25}, or trajectory-based distances~\cite{trajSimilarity:tao:2021}, \revision{at the cost of a stable output orientation, as MDS embeddings are rotationally invariant.}
For the remainder of this paper, we derive the distance matrix from pairwise geodesic distances between centroids.

\subhead{Perceptually Linear Colorspace}
Projecting our pairwise distance matrix into a three-dimensional embedding allows us to map geographical references into a given colorspace.
While UMAP or t-SNE can preserve local neighborhood structures, MDS aims to retain pairwise dissimilarities between data points, with embedding distances as close as possible to the original distances.
To satisfy distance preservation~\GoalA, we map the MDS output to the $J^{*}$, $a^{*}$, $b^{*}$ components of CAM16-UCS~\cite{li:cam16:2017}, which outperforms other perceptually uniform color spaces for both small and large color differences~\cite{luo:ucsSurvey:2023}.
The rotational invariance of MDS produces different point cloud orientations, but perceptual linearity is preserved regardless of rotation.

\subhead{sRGB Gamut Optimization}
UCSs model human perception and exceed the displayable range of standard monitors. 
The sRGB gamut, represented by a bounded polyhedron in CAM16-UCS, is a volume that encompasses colors most standard monitors can display, and serves as our main target. However, this target can be customized to any known device gamut.
We provide two optimization strategies to fit the embedding within the gamut while maximizing colorspace usage~\GoalB.

\newcommand{\gamutWrapFig}[1]{
\begin{wrapfigure}[5]{l}{0.06\textwidth}
  \vspace{-2.5em}
  \begin{center}
    \includegraphics[width=0.1\textwidth]{#1}
  \end{center}
\end{wrapfigure}
}

\begin{wrapfigure}[5]{l}{0.07\textwidth}
  \vspace{-2.5em}
  \begin{center}
    \includegraphics[width=0.1\textwidth]{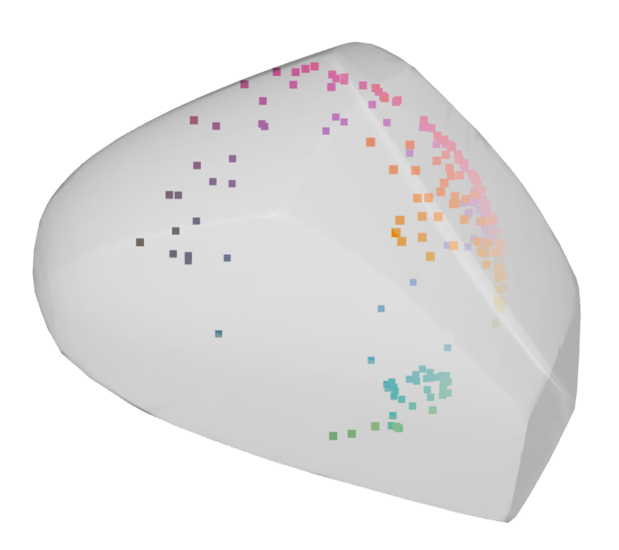}
  \end{center}
\end{wrapfigure}
\subhead{\mapSimplex} Koutrouli et al.~\cite{koutrouli:UCIE:2022} compute a convex hull of the 3D embedding and use Nelder-Mead simplex optimization to fit it inside the target gamut.
Ensuring the gamut fully contains the embedding constitutes a non-continuous constraint, making gradient-based approaches unreliable and requiring a derivative-free approach. 
We adopt this approach (see \autoref{sphericolor:fig:teaser}a) by optimizing a rigid transformation with uniform scaling that balances three objectives: preserving pairwise distances, penalizing out-of-gamut points, and rewarding expansion within gamut boundaries.
This guides the optimization to expand the embedding as much as possible without violating gamut limits.
While this preserves relative geometry and satisfies~\GoalA, colorspace usage is limited to the hull-gamut intersection~\GoalB, and the method requires hyperparameter tuning and can converge slowly.

\begin{wrapfigure}[5]{l}{0.06\textwidth}
    \vspace{-2.5em}
  \begin{center}
    \includegraphics[width=0.1\textwidth]{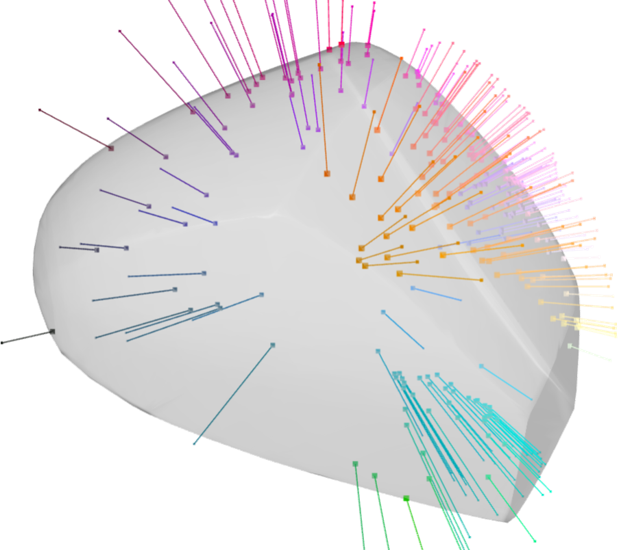}
  \end{center}
\end{wrapfigure}
\subhead{\mapRay} For better gamut exploitation, we align the embedding with the gamut centroid and project each point onto the gamut surface along rays emanating from the centroid (see \autoref{sphericolor:fig:teaser}b).
For each point, we compute the ray direction from the centroid and perform ray-triangle intersection against the gamut's triangulated hull to reposition it onto the gamut surface.
This method is parameter-free and runs in linear time with respect to the number of points. Since it interpolates the embedding from a spherical layout to the shape of the gamut, it inevitably introduces distortions to the pairwise distances, violating~\GoalA.
However, the overall distances between points increase, resulting in larger colorspace usage and greater perceptual contrast among the resulting colors~\GoalB.

\begin{figure*}[t]
    \begin{subfigure}[t]{0.24\linewidth}
        \centering
        \includegraphics[width=\linewidth]{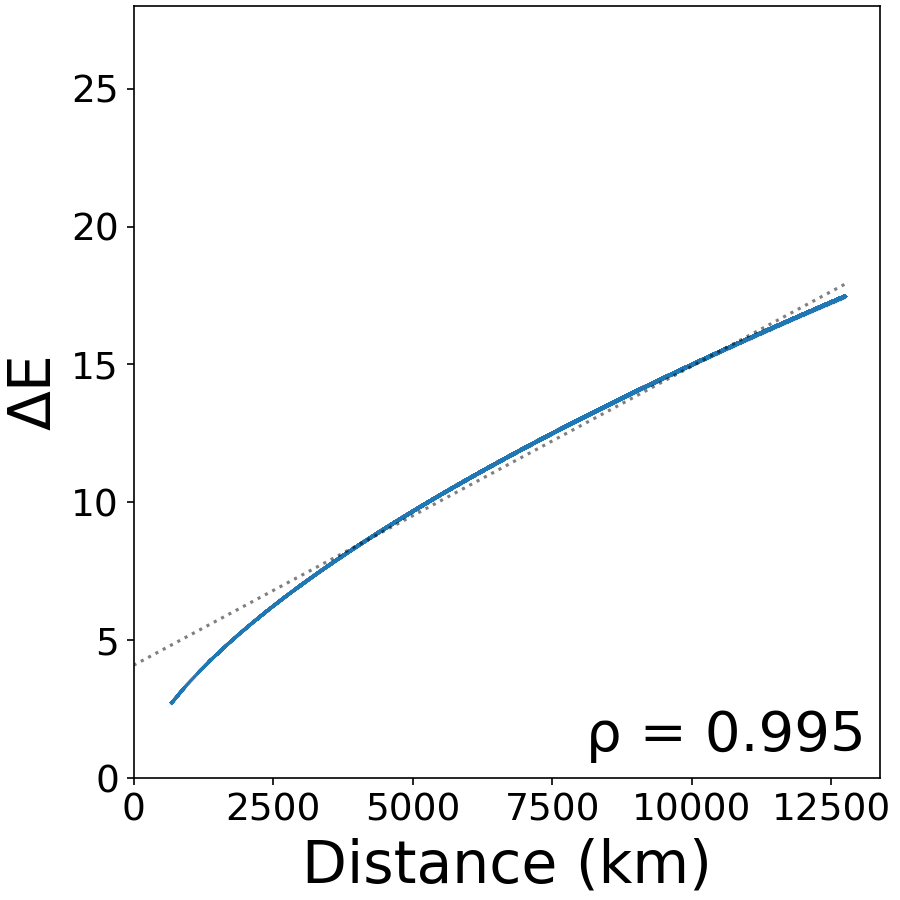}
        \caption{\mapSimplex}
    \end{subfigure}%
    \hfill
    \begin{subfigure}[t]{0.24\linewidth}
        \centering
        \includegraphics[width=\linewidth]{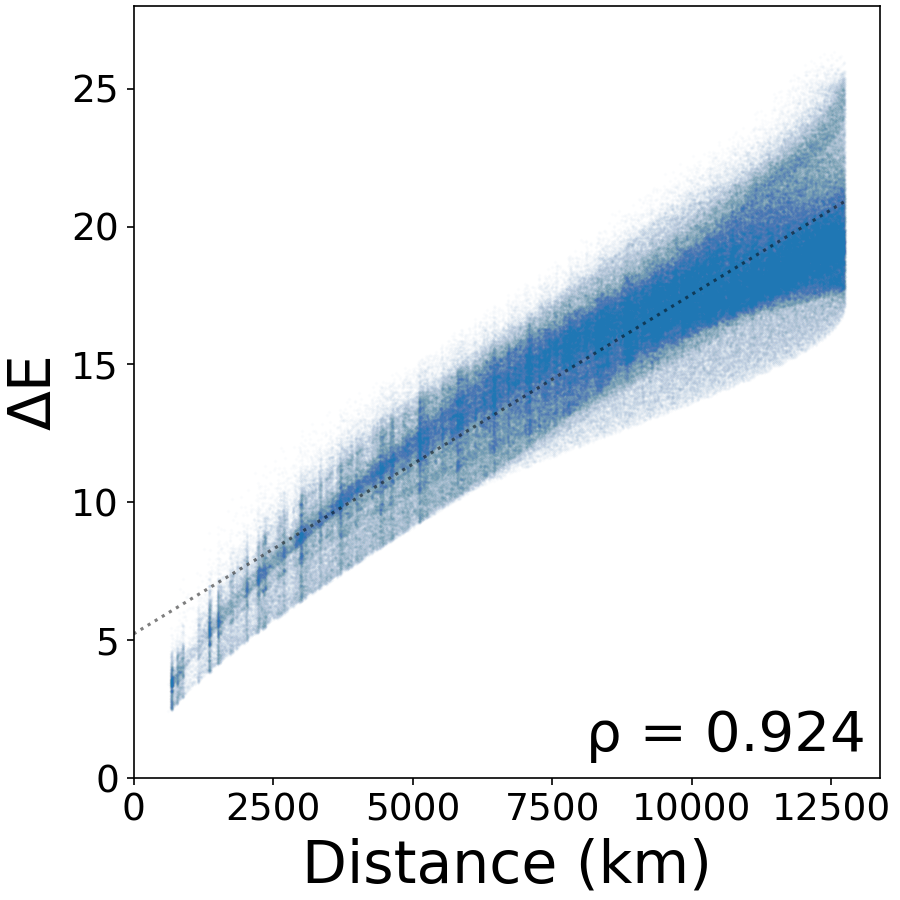}
        \caption{\mapRay}
    \end{subfigure}%
    \hfill
    \begin{subfigure}[t]{0.24\linewidth}
        \centering
        \includegraphics[width=\linewidth]{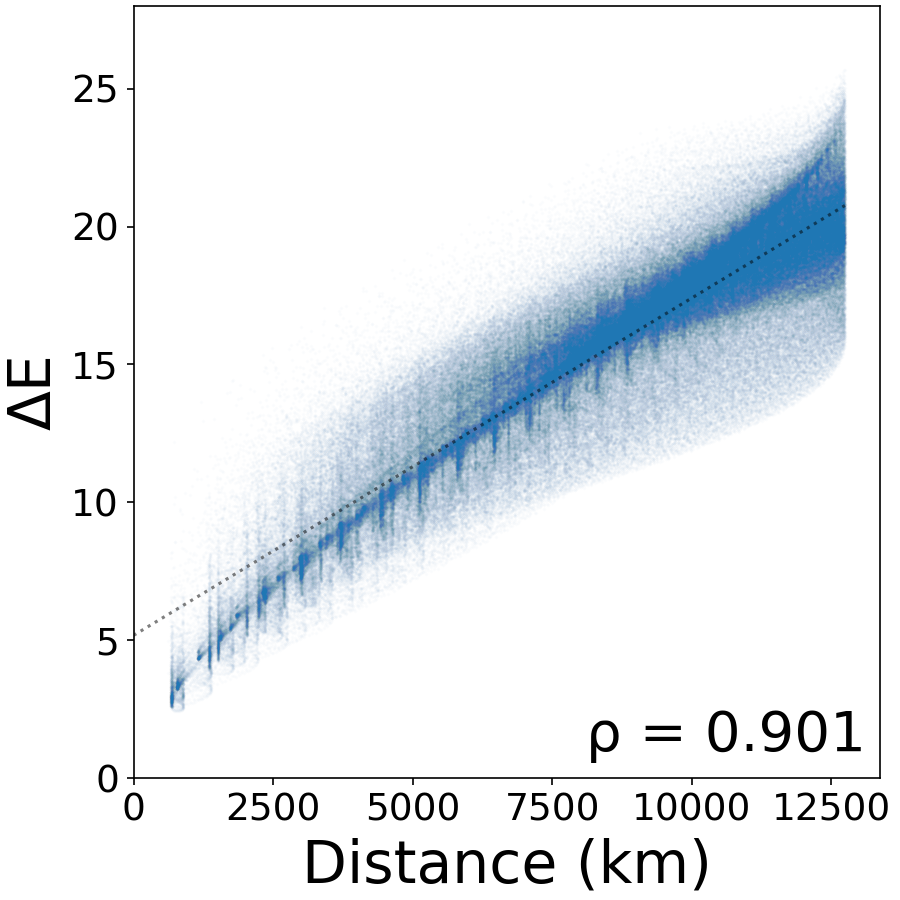}
        \caption{ HSLuv Double Cone}
    \end{subfigure}%
    \hfill
    \begin{subfigure}[t]{0.24\linewidth}
        \centering
        \includegraphics[width=\linewidth]{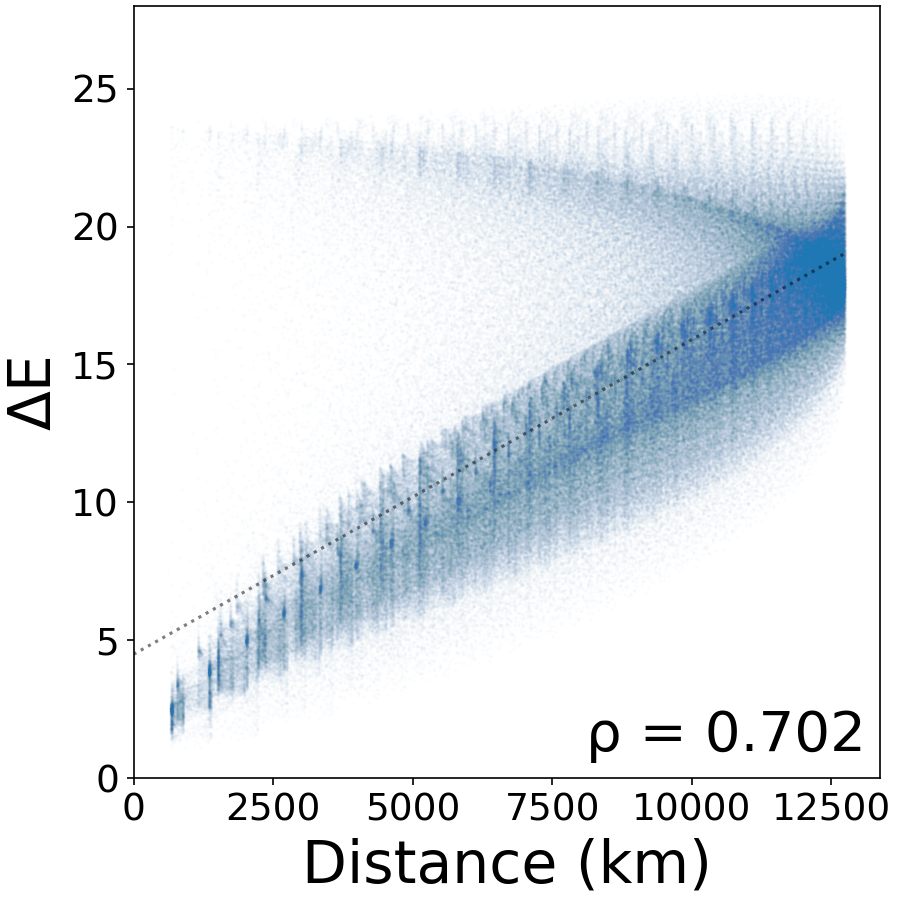}
        \caption{2D Colormap}
    \end{subfigure}
    \caption{Shepard Diagrams for the \sampleds dataset. \mapSimplex shows optimal preservation (but also the lowest maximum \revise{$\Delta E$}), whereas nonlinearities are apparent in the 2D colormap. }
    \label{sphericolor:fig:shepards}
    \vspace{\figVSpace}
\end{figure*}

\smallskip

\revise{Considering runtime, both \mapRay and \mapSimplex share an initial MDS embedding step at $O(n^2)$. \mapRay then performs a single-pass ray projection onto the gamut hull at $O(n \cdot T)$ for $T$ hull triangles, and is deterministic and non-iterative. \mapSimplex instead runs a Nelder-Mead optimization that re-evaluates pairwise stress and an out-of-gamut penalty at every step, yielding $O(k \cdot n^2)$ over $k$ iterations and requiring careful hyperparameter selection, which leads to variable and typically longer runtimes. The naïve HSLuv double cone and the 2D colormap, in contrast, are closed-form per-entity lookups in $O(n)$.}

%% file: sections/04-quantiative-eval.tex
\section{Quantitative Evaluation}
\label{sphericolor:sec:quant-eval}
We outline quantitative methods for assessing colormap quality relative to our design goals, then evaluate both naïve approaches and our methods on synthetic and real-world datasets.
As both optimizations can converge to local minima, we evaluate 20 different initial MDS rotations and select the configuration achieving the lowest objective value.

\subsection{Measures}
Bujack et al.~\cite{colormapTaxonomy:bujack:2018} provide measures for \emph{Uniformity} and \emph{Discriminative power} based on colormap speed derived from the rate of change between input and color distance. However, their framework assumes continuous 1D scalar-to-color mappings with natural ordering, which is not applicable to our spherical input topology.
The quality assessments from Bernard et al.~\cite{survey2dcolormap:bernard:15} measure properties of the static planar 2D colormap image itself rather than the relationship between input and output distances.
We therefore develop analogous measures tailored to spherical geospatial input data.


 
\subhead{\GoalA Distance Preservation}
We quantify the agreement between geodesic and perceptual color distances using Pearson correlations.
For a visual inspection, we provide a Shepard Diagram with geographic distances along the x-axis and color distances on the y-axis \revision{(see \autoref{sphericolor:fig:shepards})}.
For a local assessment, we reduce the set of distance pairs to only consider distances of K-Nearest Neighbors. 
\autoref{sphericolor:fig:local-correlation} shows the correlations as a function of the neighborhood size $k$.




\subhead{\GoalB Colorspace Exploitation}
We approximate colorspace exploitation as the fraction of sRGB gamut volume filled by the colormap's convex hull in CAM16-UCS, yielding a score in $[0,1]$ where $1$ represents complete gamut filling (see \autoref{sphericolor:fig:teaser}).
Additionally, we report the min, median, and max color difference \revise{as CAM16-UCS color differences ($\Delta E$)} to summarize the perceptual distance spread.
\revise{Under controlled viewing conditions, a $\Delta E$ above roughly 1 corresponds to about one Just Noticeable Difference (JND), suggesting perceptibly distinguishable colors.}

\subsection{Data}
The performance of a given colormap is highly dependent on its underlying data, hence we evaluate on synthetic and real-world data: \sampleds contains 1000 uniformly sampled points on a sphere \revise{($\approx$623\,km spatial resolution on Earth)} using Fibonacci Lattice; \countryds contains 212 centroids of World Bank Economic regions~\cite{WorldDevelopmentIndicators:worldbank:2014}.



\subsection{Results}
\begin{figure}[t!]
    \centering
    \includegraphics[width=\linewidth]{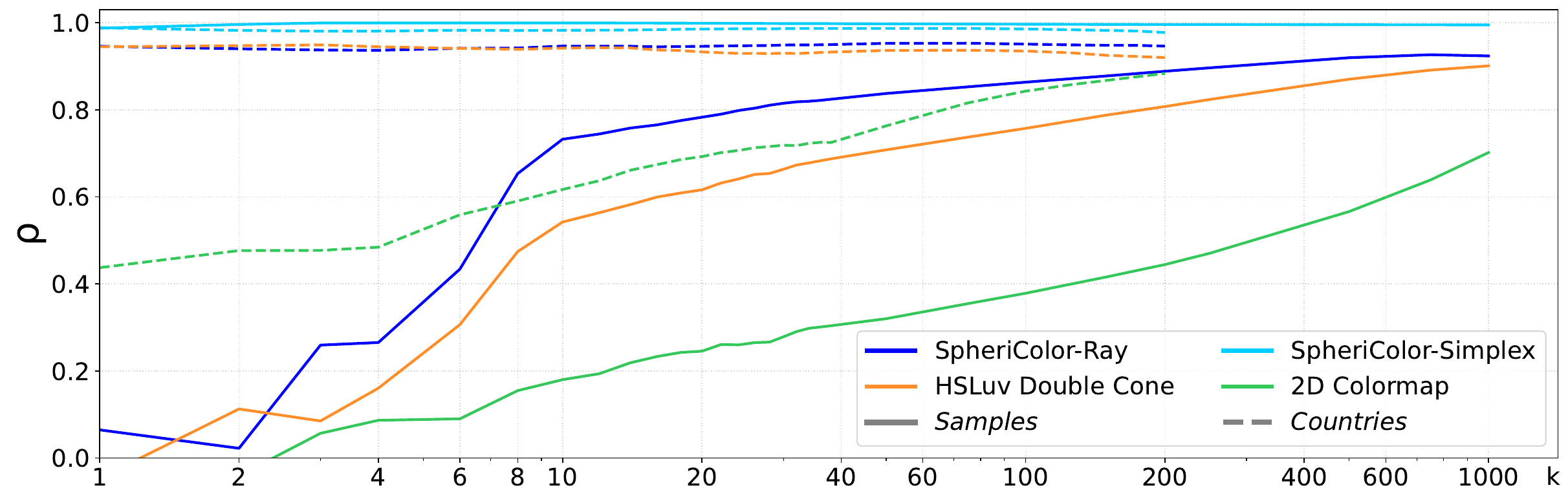}
    
    \caption{Pearson correlation values $\rho$ for different neighborhood sizes $k$ for the \sampleds dataset. Especially for smaller neighborhoods, \mapSimplex shows the highest values, indicating good local distance preservation~\GoalA. }
    \label{sphericolor:fig:local-correlation}
    \vspace{\figVSpace}
\end{figure}

\subhead{Distance Preservation~\GoalA}
\autoref{sphericolor:tab:measures} shows that \mapname approaches achieve the highest Pearson correlations across both datasets, indicating superior distance preservation.
\revision{\autoref{sphericolor:fig:shepards} confirms this, showing clear correlation patterns with a slightly bent curve, attributable to the geodesic input distances. }
\mapSimplex produces minimal distortions, while \mapRay achieves slightly lower but still strong correlations (>.89 for both datasets).
When looking at local distances (see \autoref{sphericolor:fig:local-correlation}), \mapSimplex outperforms all other approaches. 
\revision{If only the nearest-neighbor distances are considered ($k=1$), it maintains its high score for \sampleds (.98), while all other approaches do not exhibit any correlation with scores around $0$.
With increasing neighborhood $k$, other approaches also show increased correlation.
\mapRay reaches a strong positive correlation (>.7) with $k=9$, while HSLuv, the next best, achieves a strong correlation with $k=50$.
This shows that \mapSimplex is able to represent the geographical input distances most faithfully, also on a local scale.}
The HSLuv double cone achieves correlations around .90 for both datasets, yielding slightly worse results compared to \mapRay.
The 2D colormap performs adequately for \countryds (.88) but considerably worse for \sampleds (.67).
This can be traced back to distances crossing the projection's antimeridian, which produce large color differences for small geodesic distances (see \autoref{sphericolor:fig:shepards}d).

\subhead{Colorspace Exploitation~\GoalB} 
On \countryds, \mapRay outperforms all approaches (.82 gamut intersection) with the highest minimum (0.77 \revise{$\Delta E$}) and maximum (25.35 \revise{$\Delta E$}) color distances.
\mapSimplex spans only .41 of the gamut with the lowest maximum color distance (19.92 \revise{$\Delta E$}).
The 2D colormap achieves the lowest exploitation (.07). Planar colormaps model a curved slice through the colorspace rather than a volumetric region, which is clearly visible in \autoref{sphericolor:fig:teaser}c.
\revision{Gamut intersection naturally increases with the number and uniformity of input points, as denser coverage leaves fewer unoccupied gamut regions.}
\mapRay (.99) and HSLuv double cone (.97) hence nearly fill the sRGB gamut for \sampleds, whereas
\revision{\mapSimplex (.41) and 2D colormap (.21) provide substantially lower volume intersections.}

Considering both goals, \mapRay arguably provides the best tradeoff with strong performance on both \GoalA and \GoalB.
While \mapSimplex better preserves distances, its limited gamut usage produces less distinguishable, less saturated colors, evidenced by lower gamut intersection and \revise{$\Delta E$} scores.

%% file: sections/06_discussion.tex
\section{Discussion}
The results show that \mapname outperforms the evaluated naïve approaches with respect to distance preservation \GoalA and colorspace exploitation \GoalB,
\revise{ making it well suited to color PCPs or scatterplots, revealing geographic clusters while freeing position for other attributes.}
However, both objectives cannot be optimized simultaneously, as improving one typically compromises the other.
We propose the following guidelines for constructing colormaps in spherical geospatial visualizations:

\textbf{When \emph{distance preservation} is prioritized, \mapSimplex is preferable.}
Rigid transformations preserve local distances most faithfully, at the cost of reduced colorspace utilization and saturation.

\textbf{When \emph{colorspace exploitation} is prioritized, \mapRay is the better choice. }
It tends to produce more distinguishable and saturated colors, though distance preservation may be reduced.

Performance further depends on the embedding geometry and the target gamut. \mapRay is specifically designed for spherical embeddings, whereas \mapSimplex is more adaptable and may better fit planar or irregular embeddings inside the gamut.
\vspace{1em}

\subhead{Discriminability} 
\revise{On \countryds, all approaches exhibit a minimum color distance below $1\,\Delta E$ (see \autoref{sphericolor:tab:measures}), caused by very small geodesic distances between neighboring countries.
\mapSimplex still achieves the largest minimum color distance among the distance-preserving methods. 
For \sampleds, increasing to 5000 points ($\approx$280\,km spatial resolution) still remains above $1.5\,\Delta E$, a comfortable margin over the $1\,\Delta E$ JND.
The gamut-filling variants further vary lightness more strongly, which aids discriminability for large choropleth areas but may reduce legibility for small marks such as scatterplot points.}

\newcommand{\rup}{{\small($\uparrow$)}\xspace}
\begin{table}[t!]
    \centering
    \setlength{\tabcolsep}{2pt}
    \resizebox{\linewidth}{!}{%
    \begin{tabular}{@{}|c|c| c c c | c c c c | @{}}
    \hline
        \multirow{3}{*}{\rotatebox{90}{\textbf{Data}}} & \multirow{3}{*}{\textbf{Colormap}} & \multicolumn{3}{c|}{\textbf{\GoalA}} & \multicolumn{4}{c|}{\textbf{\GoalB}} \\
         & & \multicolumn{3}{c|}{Pearson $\rho$~\rup} & \multicolumn{3}{c}{\revise{$\Delta E$} \rup} & Gamut  \\
         & & $k=1$ & $k=n/2$ & $k=n$ & Min & Med & Max & Intersect \rup \\ \hline
        \multirow{4}{*}{\rotatebox{90}{\countryds}} & \mapSimplex & \textbf{.9885} & \textbf{.9857} & \textbf{.9775} & .63 & 12.59 & 19.92 & .4186 \\
        & \mapRay & .9460 & .9509  & .9462 & \textbf{.77} & \textbf{14.99} & \textbf{25.35} & \textbf{.8294} \\
         & HSLuv Double Cone & .9447 & .9350 & .9200 & .72 & 13.74 & 23.17 & .6346 \\
         & 2D Colormap  & .4374 & .8428 & .8835 & .00 & 11.39 & 23.68 & .0748 \\  \hline
        \multirow{4}{*}{\rotatebox{90}{\sampleds}} & \mapSimplex & \textbf{.9894} & \textbf{.9903} & \textbf{.9639} & \textbf{2.61} & 14.05 & 17.49 & .4127 \\
        & \mapRay & .0379 & .9143 & .8934 & 2.48 & \textbf{16.37} & \textbf{26.29} & \textbf{.9902} \\
         & HSLuv Double Cone & -.046 & .8655 & .8715 & 2.43 & 16.28 & 25.68 & .9751 \\
         & 2D Colormap  & -.096 & .5631 & .6777 & 1.08 & 14.52 & 24.80 & .2130 \\ 
         \hline
    \end{tabular}
    }
    \caption{Quality measures to assess the coverage of the design goals for both a real-world and a synthetic dataset. The best performing approach is highlighted in bold for every measure and dataset.}
    \label{sphericolor:tab:measures}
    \vspace{\figVSpace}
\end{table}


\subsection{Limitations and Future Work}

\subhead{Evaluation}
Our evaluation is purely quantitative. In absence of existing quality measures for spherical input data, we derived metrics for distance preservation and colorspace exploitation based on established design rules~\cite{survey2dcolormap:bernard:15, colorgeometry:bujack:2025}.
\revision{While distance preservation builds on standard measures, our gamut-intersection score is a reasonable proxy for~\GoalB but not an established quality measure.}
A user study is needed to validate whether these metrics align with human perception and task performance in real-world geospatial visualizations.

\subhead{Stability}
In contrast to the naïve approaches, MDS embeddings are rotationally invariant and sensitive to the point configuration, so adding a single entity can yield a substantially different colormap.
Replacing MDS with a canonical spherical-to-3D transformation aligned to the $J^*$ lightness axis could address this for \mapRay.
This additionally gives semantic meaning to the colors as latitude is encoded through lightness, at the cost of reduced colorspace exploitation~\GoalB.
\revision{Applications requiring a fully dataset-independent mapping may instead use the HSLuv double cone, which is inferior in terms of the design goals but covers the full spherical domain by design.}

\subhead{Generalizability}
While we demonstrated the capabilities of \mapname using simple point-based geodesic distances, the distance matrix could be adapted to model more complex geospatial similarities.
The resulting 3D MDS embedding may no longer approximate a sphere, in which case \mapRay could collapse points with the same angular direction onto the same gamut location. 

%% file: sections/07_conclusion.tex
\section{Conclusion}
\label{sphericolor:sec:conclusion}

We introduce \mapname, a novel approach for mapping spherically bound geospatial references to color. 
By leveraging a geospatial distance function and multidimensional scaling, we project the entities into a perceptually linear colorspace and propose two gamut-constrained optimization techniques to ensure the preservation of geodesic distances and the maximization of the available colorspace.
Through a quantitative evaluation using real-world and synthetic datasets, we demonstrate the performance of these optimizations and compare them to a 2D colormap and a HSLuv double cone.
We show that \mapSimplex best preserves geodesic distances, while \mapRay optimizes colorspace exploitation.